\documentclass[english, aip, cha, reprint]{revtex4-1}
\usepackage[T1]{fontenc}
\usepackage[utf8]{inputenc}
\usepackage{amstext}
\usepackage{amssymb}
\usepackage{graphicx}
\usepackage{babel}
\usepackage[caption=false]{subfig}
\usepackage{color}
\begin{document}

\title{Chimera-like States in Structured Heterogeneous Networks}

\author{Bo Li}
\email{bliaf@connect.ust.hk}
\affiliation{Department of Physics, The Hong Kong University of Science and Technology, Hong Kong}
\author{David Saad}
\affiliation{Non-linearity and Complexity Research Group, Aston University, Birmingham B4 7ET, United Kingdom}
\date{\today}

\begin{abstract}
Chimera-like states are manifested through the coexistence of synchronous and asynchronous dynamics and have been observed in various systems. 
To analyze the role of network topology in giving rise to chimera-like states we study a heterogeneous network model comprising two group of nodes, of high and low degrees of connectivity.
The architecture facilitates the analysis of the system, which separates into a densely-connected coherent group of nodes, perturbed by their sparsely-connected drifting neighbors. 
It describes a synchronous behavior of the densely-connected group and scaling properties of the induced perturbations.
\end{abstract}
\maketitle

\begin{quotation}
Nonlinear interactions of coupled oscillators, such as the Kuramoto model with sinusoidal couplings, give rise to fascinating collective behaviors~\citep{Acebron2005,Boccaletti2002,Rodrigues2016}, one of which is the emergence of chimera states in identical and symmetric networks, namely, the coexistence of stable synchronous and fluctuating asynchronous dynamical patterns~\citep{Panaggio2015}. 
Studies have shown that the chimera-like states also occur in complex networks, which exhibit disparate behaviors for nodes of different connectivities, for instance higher-degree nodes are more prone to synchronization than lower-degree nodes in scale free networks~\citep{Zhu2014}. 
However, the rich structure of complex networks makes it difficult to analyze the dynamics and gain insight into the collective dynamics and its properties. 
To facilitate the analysis and better understand the role of network heterogeneity, we examine a structured heterogeneous network model comprising only two interacting groups of nodes, one of which is densely connected while the other is sparsely connected. 
As in other complex networks this model exhibits a chimera  state-like behavior whereby high-degree nodes form a synchronized domain while low-degree nodes remain unsynchronized. 
The main benefit of this topology is that it is amenable to analysis and provides insight that is difficult to obtain otherwise; we derive a self-consistent mean field theory for these networks, which explains quantitatively their behavior, especially the synchronized dynamics of the densely connected component under perturbation induced by the sparsely connected nodes.
\end{quotation}

\section{Introduction}

The dynamics of coupled oscillators has been widely studied to understand the synchronization behaviors in complex systems~\citep{Acebron2005,Boccaletti2002,Rodrigues2016}.
Among the fascinating phenomena observed in these systems the chimera states, which exhibit a coexistence of synchronous and asynchronous domains, have attracted particular interest recently~\citep{Panaggio2015}. It came as a surprise that the symmetry-breaking chimera states were observed in symmetric and homogeneous media~\citep{Kuramoto2002,Shima2004,Abrams2004,Abrams2008,Ashwin2015,Bick2017}.
This phenomena is also reminiscent of the coexistence of equilibrium-like domains in non-equilibrium systems~\citep{Saad2013}.

The analysis of chimera states is generally difficult, due to the heterogeneous interactions between a large number of variables, and appropriate approximations are required. One of the approximations is based on mapping the system onto a low dimensional manifold~\citep{Abrams2008}, which facilitates a simplification that leads to tractable solutions.
Such approaches are either based on the continuous self-consistent mean field theory or rely on the assumption of Ott-Antonsen reduction to simplify the problem~\citep{Kuramoto2002,Ott2008}. 
Carefully chosen initial conditions are also necessary to produce chimera states in these symmetric systems~\citep{Martens2016}.
In addition to the theoretical discovery, chimera states have also been observed and studied in different experiments~\citep{Tinsley2012, Hagerstrom2012, Martens2013, Kapitaniak2014}. 
  
While it is fascinating to understand the emergence of chimera states as symmetry-broken states in very large homogeneous symmetric networks, real systems are typically heterogeneous and of finite size~\citep{Strogatz2001, Albert2002}. To explore synchronization behavior in real systems, one has to take into account their complex and heterogeneous network structure~\citep{Arenas2008, Rodrigues2016}. Two recent examples are the modelling of neural activities in the resting-state functional networks~\citep{Vuksanovic2014} and the modular neural networks of C. Elegans~\citep{Hizanidis2016}, both of which consider irregular network topologies. Studies of networks comprising non-identical constituents with heterogeneous natural frequencies or coupling strengths, and irregular topologies have shown similar phenomena where synchronized and unsynchronized subpopulations coexist~\citep{Ko2008, Laing2009a, Laing2009b, Laing2012, Zhu2014, Omelchenko2015, Buscarino2015, Jiang2016, Olmi2016}. Some of the features exhibited are different from those of symmetric systems, e.g., in the case of all-to-all coupled oscillators with scale-free distributed coupling strengths, the phase locked subpopulations are the oscillators with small coupling strength~\citep{Ko2008, Laing2009b}.  Another study that does not rely on the system's symmetry~\citep{Zhu2014} investigated sinusoidally coupled oscillators in Erd{\"o}s-R{\'e}nyi  and scale-free networks, which exhibit chimera-like state behavior from arbitrary initial conditions.

An interesting observation from studies in scale free networks is that nodes of high degrees are more likely to synchronize than nodes of low degrees~\citep{Zhu2014}. 
This is an example where heterogeneity in network topology results in disparate behaviors across the system~\citep{Zhou2006,Gardenes2007,Gardenes2011}.
Also in models with chaotic oscillators, it was shown that hubs synchronize while low-degree nodes do not~\citep{Zhou2006,Pereira2010}. 
It is proposed in a theoretical study of neural culture that neurons with more inputs, i.e., with high in-degrees, are the leaders of the burst activities~\citep{Eckmann2010}. 
In general, the strong interactions that the highly-connected oscillators have with their neighbors distinguish them from the non-synchronizing low-degree nodes~\citep{Zhu2014}.
However, the complex structure of heterogeneous networks, such as scale
free or small world networks is not amenable to analysis due to the combined complexity of the network topology and intractability of the large scale-dynamical model. In this work, we investigate a simple heterogeneous
network architecture which supports an easy-to-reach chimera-like state but is amenable to analysis. It exposes the synchronized equilibrium-like behavior of the densely-connected components and facilitate the derivation of perturbation originated by the sparsely-connected components; finite-size scaling of these perturbation and the corresponding order parameter is also be derived, shedding light on the interplay between the two types of nodes.

\section{The Model}

We consider a heterogeneous network with degree distribution of two
peaks, i.e., the network comprises two classes of nodes, one
of which has extensive connectivity scaled with
system size $N$ while the nodes in the other has finite connectivity.
Denote the two classes as the dense group $\mathcal{D}$ and the sparse
group $\mathcal{S}$ respectively. By assigning a large number of intra-group
connections between constituents of the dense group $\mathcal{D}$, each node
inside $\mathcal{D}$ can receive large number of coordinated signals
from within the group, dominating the fluctuations of other neighboring nodes,
potentially leading to coherent intra-group dynamics. Conversely,
the nodes in the sparse group can be easily perturbed by fluctuation induced by any of its neighbors
due to their limited connectivity.

\begin{figure}
\includegraphics[scale=0.4]{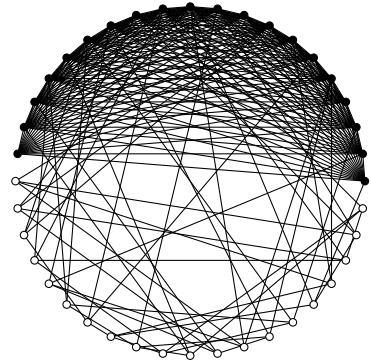}

\caption{An instance of the heterogeneous network composed of a densely connected
 (black nodes) and sparsely connected groups (white nodes), where
$N=40$, $d_{\small{\cal S}}=4$. \label{fig:network_nn40}}
\end{figure}

Specifically, to simplify the construction, each node $i\in\mathcal{D}$ is connected to $d_{\small{\cal D}}(\propto N)$
other nodes randomly chosen from the same class $\mathcal{D}$, and
each node $j\in\mathcal{S}$ is connected to $d_{\small{\cal S}}(\sim O(1))$ other
nodes randomly chosen from the whole system. Let $n_{\small{\cal D}}$ and $n_{\small{\cal S}}$
denote the number of nodes in the dense and sparse groups, respectively.
For simplicity, we consider the special case where node $i\in\mathcal{D}$
is connected to all the rest members of $\mathcal{D}$ so that the
dense group forms a complete subgraph, and both groups include the same number
of nodes, i.e., $d_{\small{\cal D}}=n_{\small{\cal D}}-1$ and $n_{\small{\cal D}}=n_{\small{\cal S}}=N/2$. Fig.~\ref{fig:network_nn40}
is an instance of the proposed network structure with $N=40$ and
$d_{\small{\cal S}}=4$.
We would like to remark that the chimera-like state discussed below is not restricted to the special case of the dense group being a complete subgraph and $n_{\small{\cal D}} = n_{\small{\cal S}}$, but is a more general phenomena in the networks of a similar construction to that described above. Nevertheless, the simplification makes the analysis easier.

The governing dynamics we consider is the Kuramoto model with phase lag in the chimera state studies~\citep{Kuramoto2002,Shima2004,Abrams2008}
\begin{eqnarray}
\dot{\theta}_{i} & = & \omega+\frac{K}{n_{\small{\cal D}}}\sum_{j\in\partial i}\sin(\theta_{j}-\theta_{i}-\alpha)\nonumber \\
 & = & \omega+\frac{K}{n_{\small{\cal D}}}\sum_{j\in\partial i\cap\mathcal{D}}\sin(\theta_{j}-\theta_{i}-\alpha)\nonumber \\
 &  & +\frac{K}{n_{\small{\cal D}}}\sum_{j\in\partial i\cap\mathcal{S}}\sin(\theta_{j}-\theta_{i}-\alpha),\label{eq:dynamics_def}
\end{eqnarray}
where $\theta_i$ is the phase of node $i$, $\omega, K$ and $\alpha$ are the natural frequency, coupling strength and the phase lag respectively, and $\partial i$ stands for the set of nodes adjacent to node $i$ where we have isolated the contribution from the dense and sparse members. 
Any node $i\in\mathcal{D}$ experiences the global mean field induced by the whole dense group $r_{\small{\cal D}}e^{i\psi_{\small{\cal D}}}:=1/n_{\small{\cal D}}\sum_{j\in\mathcal{D}}e^{i\theta_{j}}$ and the local field of its sparse neighbors $\rho_{i}e^{i\varphi_{i}}:=\sum_{j\in\partial i\cap\mathcal{S}}e^{i\theta_{j}}$, both terms are of order 1,
\begin{eqnarray}
\dot{\theta}_{i} & = & \omega+\frac{K}{n_{\small{\cal D}}}\sin\alpha+Kr_{\small{\cal D}}\sin(\psi_{\small{\cal D}}\sin(\psi_{\small{\cal D}}-\theta_{i}-\alpha)\nonumber \\
 &  & +\frac{K}{n_{\small{\cal D}}}\rho_{i}\sin(\varphi_{i}-\theta_{i}-\alpha).\label{eq:dynamics_dense_theta}
\end{eqnarray}
By construction, the dense group itself forms a $d_{\small{\cal D}}$ regular subgraph,
which can reach perfect phase synchronization if the contribution
from the sparse neighbors are negligible~\citep{Skardal2015}, i.e.,
$\theta_{i}=\Omega t+const$, $\Omega=\omega-K(n_{\small{\cal D}}-1)/n_{\small{\cal D}}\sin(\alpha)$,
for all $i\in\mathcal{D}$. This will hold true in the thermodynamic
limit $n_{\small{\cal D}}\to\infty$; in which case the densely-connected group ${\cal D}$ on its own exhibits an equilibrium-like behavior within the global non-equilibrium system~\cite{Saad2013}. In finite systems, the effect of sparse neighbors can be regarded as a finite size perturbation with strength of order $1/n_{\small{\cal D}}$ on
the coherent dynamics of the densely-connected members $\theta_{i}(t)=\Omega t+\varepsilon_{i}(t)$.

The argument cannot be applied to the sparse group. Firstly, we notice
that for node $j\in\mathcal{S}$, $|\dot{\theta}_{j}|\leq\omega+Kd_{\small{\cal S}}/n_{\small{\cal D}}\ll\Omega$,
thus it does not follow the coherent dynamics of the densely-connected group and
global synchronization of the whole system is not attainable. Secondly,
the irregularity of the sparse subgraph and presence of phase lag
interfere with phase synchronization. Finally, sparse variables are easily disturbed by noisy signals from their
neighbors due to their finite connectivity. We define the global order
parameter of the sparse group $r_{\small{\cal S}}:=\left|1/n_{\small{\cal S}}\sum_{i\in\mathcal{S}}e^{i\theta_{i}}\right|$
measuring their incoherence.

\section{Analysis and Results}

\begin{figure}
\subfloat{
\includegraphics[scale=0.2]{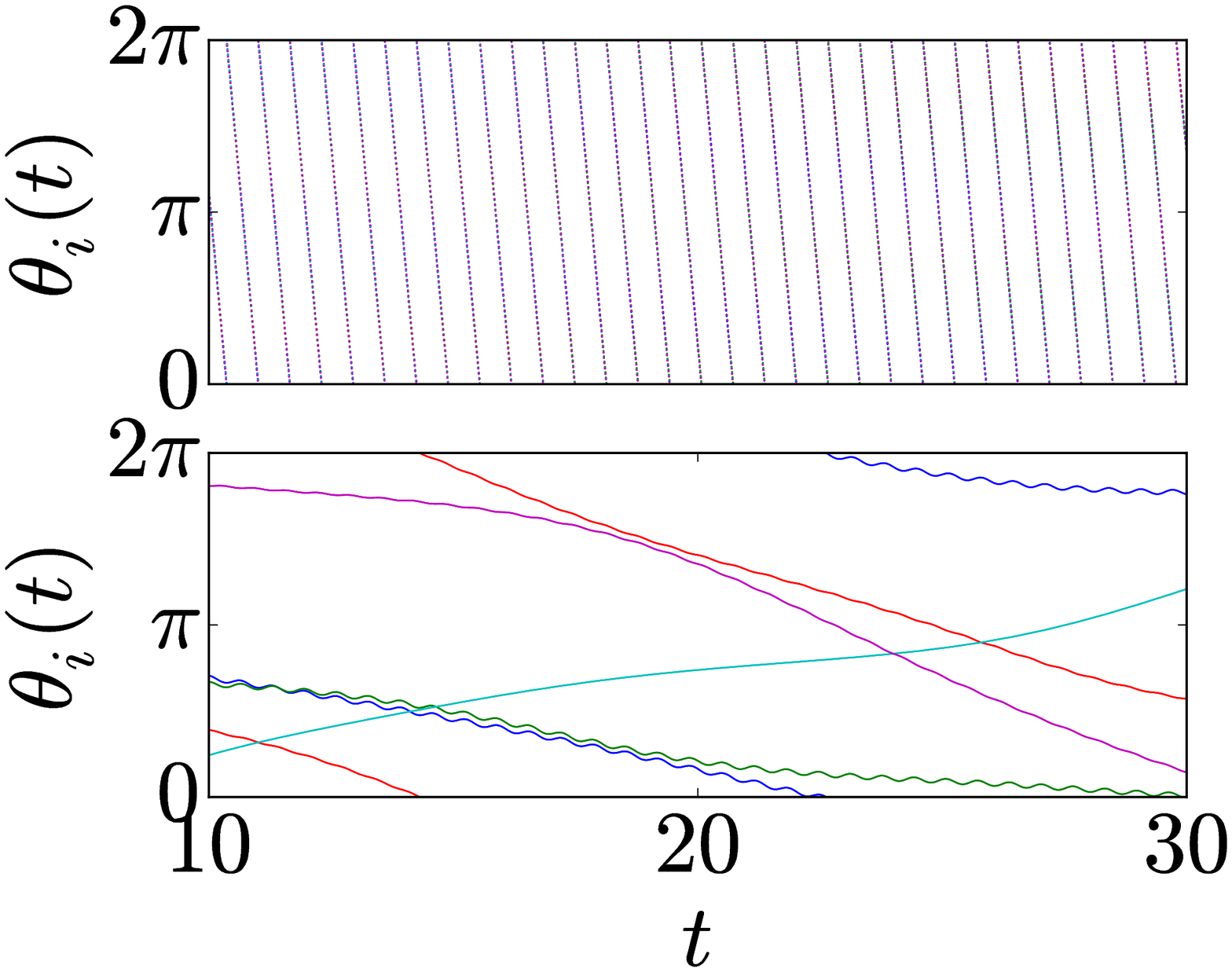}}
\subfloat{
{\includegraphics[scale=0.2]{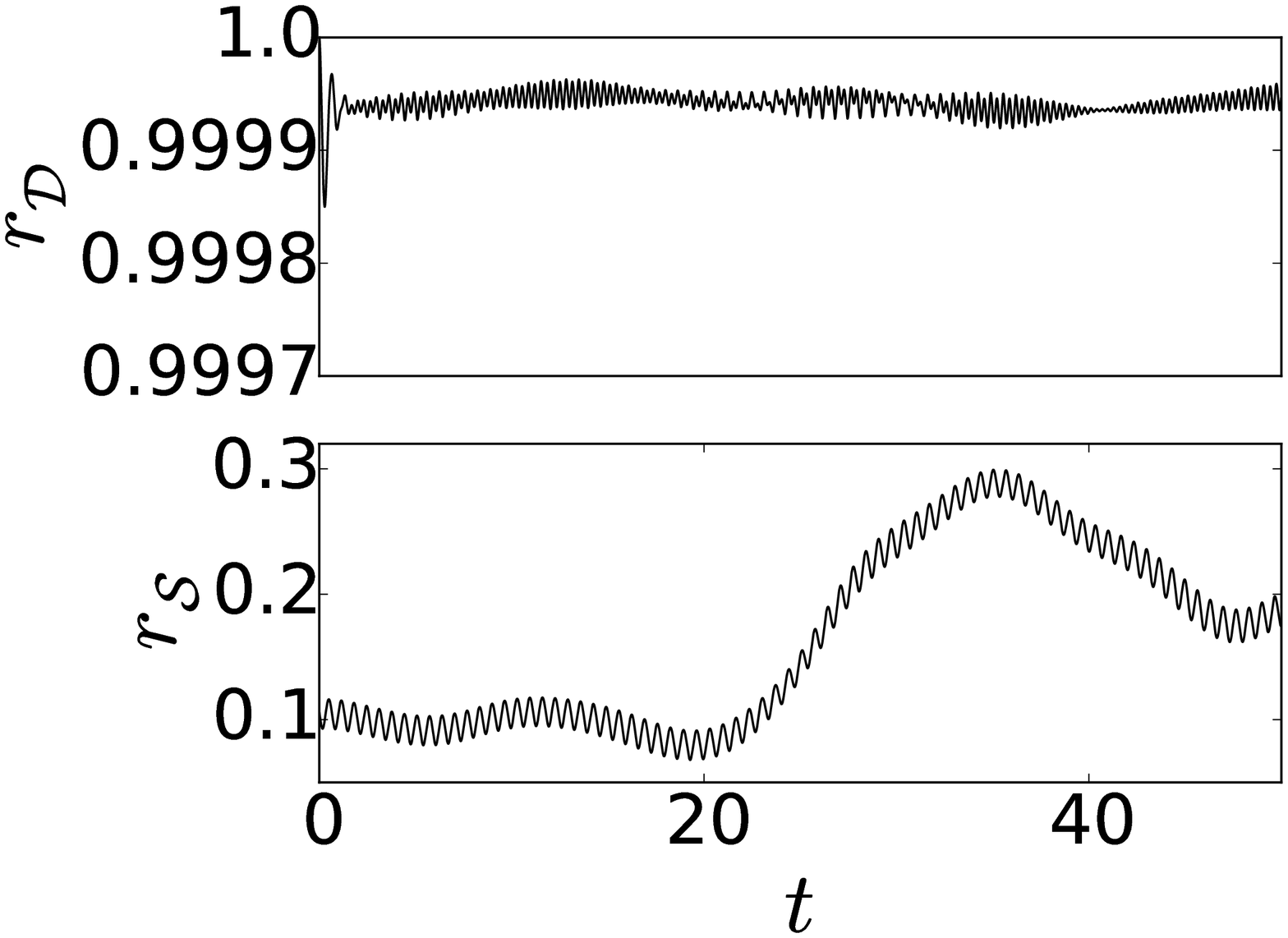}}
\begin{picture}(0,0)
\put(-45,49){\includegraphics[scale=0.052]{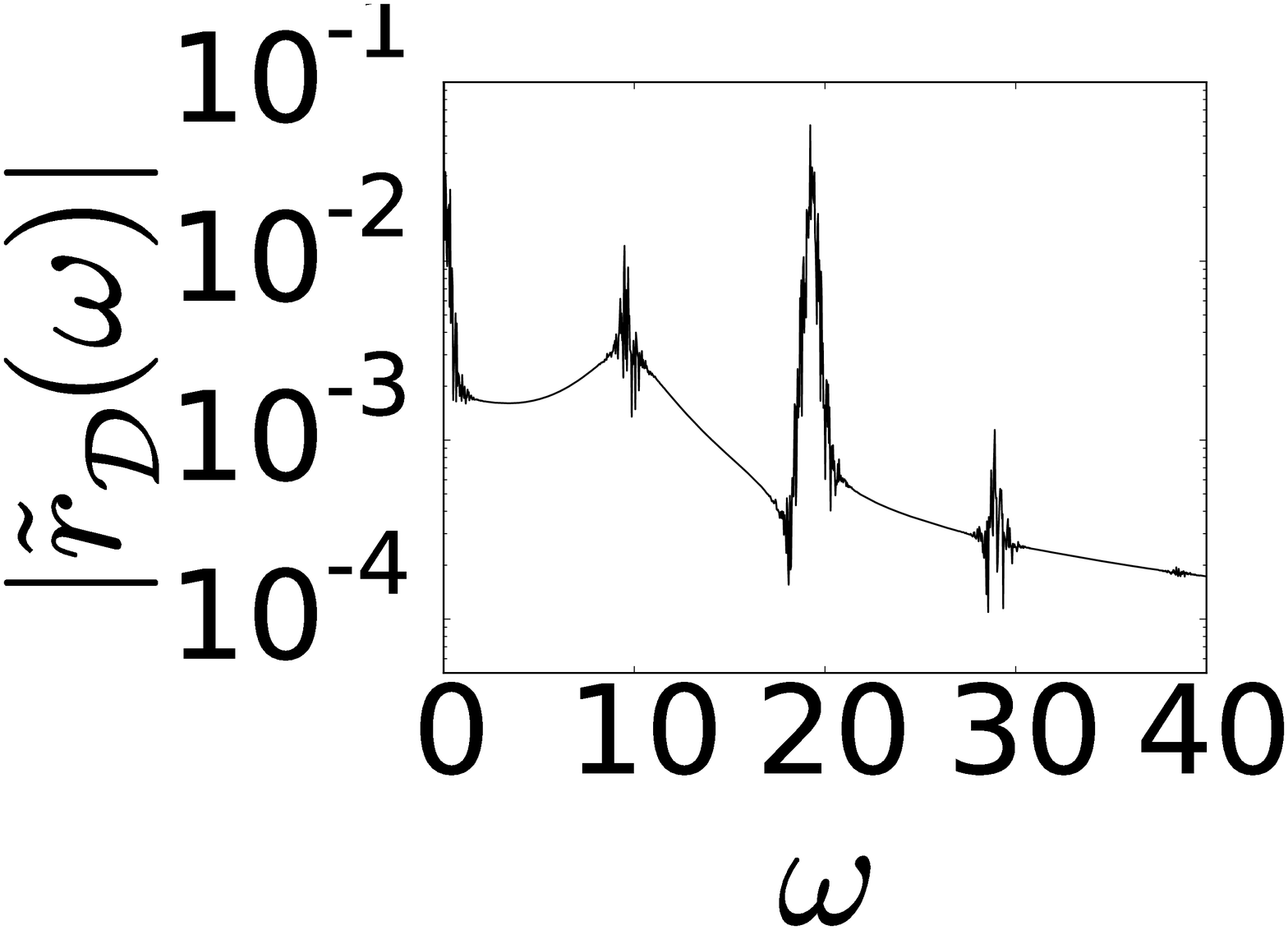}}
\end{picture}
}

\caption{(Color online) (a) Dynamical process of 5 densely (top) and
5 sparsely connected variables (bottom) in a network of size $N=200$, where the sparse degree is $d_{\small{\cal S}}=6$, coefficient $K=10$, and phase
$\alpha=\frac{\pi}{2}-0.2$. In this case, $\Omega=-K(n_{\small{\cal D}}-1)/n_{\small{\cal D}}\sin(\alpha)\approx9.7$.
(b) Evolution of the global order parameters for the densely
(top) and sparsely connected variables (bottom). Inset: Power spectrum of $r_{\small{\cal D}}(t)$,
which has its main peak at $\omega=2\Omega$.\label{fig:dynamic_demo} }

\end{figure}

\begin{figure}
\includegraphics[scale=0.4]{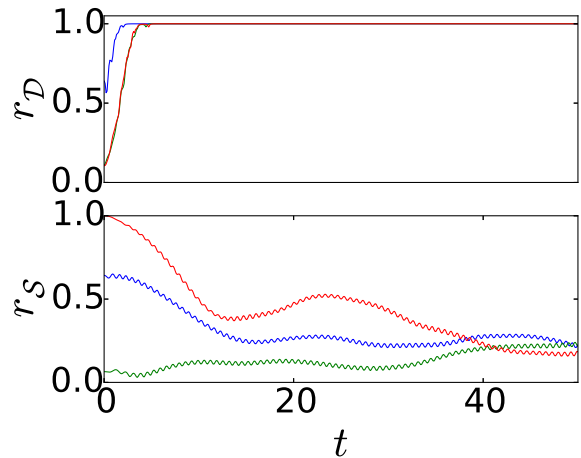}\caption{(Color online) Evolution of the global order parameters for the dense (top) and sparse (bottom) groups, starting from different
initial conditions. Blue: the initial phases of all variables are
drawn from the uniform distribution ranging from $0$ to $\pi$. Green: the initial
phases of all variables are drawn from the uniform distribution ranging from $0$
to $2\pi$. Red: the initial phases of all dense variables are drawn
from the uniform distribution ranging from $0$ to $2\pi$, while that of sparse
variables share the same phase value. \label{fig:different_init_cond} }

\end{figure}

Without loss of generality, let $\omega=0$ in the following discussion.
In Fig.~\ref{fig:dynamic_demo}, we demonstrate the dynamics of a
network with $N=200$ and $d_{\small{\cal S}}=6$ governed by Eq.~(\ref{eq:dynamics_def}).
The dense variables dynamics is coherent with a high phase velocity exhibiting minor
deviations from each other, revealed by the small deviation of $r_{\small{\cal D}}(t)$
from 1, shown in Fig.~\ref{fig:dynamic_demo}(b). On the other hand, the
sparse variables are drifting slowly and incoherently. The observed
partial synchronization phenomenon constitutes a chimera-like state. Interestingly,
the chimera-like state is easily obtained in the proposed topology
by starting with arbitrary initial conditions, as shown in Fig.~\ref{fig:different_init_cond}.
This is similar to what has been observed previously~\citep{Zhu2014}.
We remark that the chimera-like states observed here are different from the classical symmetry-breaking chimera states~\citep{Kuramoto2002,Shima2004,Abrams2004,Abrams2008,Ashwin2015,Bick2016} due to the lack of symmetry between the dense and sparse groups. 

In the following, we analyse the deviation of the densely-connected variables from perfect synchronization due to the interaction with the sparsely-connected group.
The method used is similar to the self-consistent mean field theory commonly used in the analysis of the Kuramoto model with inhomogeneous natural frequencies~\citep{kuramoto1984, Acebron2005}, in which the system is divided into two subpopulations to be determined, a phase-locked group and a drifting group, depending on the natural frequency with reference to the mean field order parameter. 
Assuming the system is in the stationary state, the two subpopulations are described by certain probability densities to be solved self-consistently to obtain the order parameter. A similar method with a space-dependent order parameter was used to analyze chimera states in finite dimensional spaces~\citep{Kuramoto2002, Shima2004}. 

In the current study, the system has homogeneous natural frequencies but highly inhomogeneous network topologies, and the phase-locked group and drifting group can be identified by the network connectivities. Instead of solving the invariant density for the whole system, we utilize the property of time scale separation in the system and isolate the synchronized group for consideration. One advantage of the chosen network topology is that the node in the dense group receives overwhelmingly more inputs from the same group than from the sparse group, allowing one to expand the deviation from perfect synchronization in orders of $1/n_{\small{\cal D}}$.  We then express the dynamics of the dense group as small deviations from perfect synchronization and solve for the small deviation self-consistently. 

Expanding the phase of a dense group variable $i$ with respect to small perturbation $\varepsilon_{i}$ such that $\theta_{i}(t)=\Omega t+\varepsilon_{i}(t)$,
and defining $g(\varepsilon)=1-r_{\small{\cal D}}(t)$, $f(\varepsilon)=\psi_{\small{\cal D}}(t)-\Omega t$ one obtains
\begin{eqnarray}
r_{\small{\cal D}}e^{i\psi_{\small{\cal D}}} & = & e^{i\Omega t}\frac{1}{n_{\small{\cal D}}}\sum_{i\in\mathcal{D}}e^{i\varepsilon_{i}} \nonumber \\
 & = & e^{i\Omega t}\left[1+\frac{i}{n_{\small{\cal D}}}\sum_{i\in\mathcal{D}}\varepsilon_{i}-\frac{1}{2n_{\small{\cal D}}}\sum_{i\in\mathcal{D}}\varepsilon_{i}^{2}+O(\varepsilon^{3})\right],\label{eq:r_D}
\end{eqnarray}
and from results in leading order in $\varepsilon^2$ (See Appendix \ref{appendix_a}),
\begin{equation}
g(\varepsilon)\approx\frac{1}{2n_{\small{\cal D}}}\sum_{i\in\mathcal{D}}\varepsilon_{i}^{2}-\frac{1}{2n_{\small{\cal D}}^{2}}\left(\sum_{i\in\mathcal{D}}\varepsilon_{i}\right)^{2},\label{eq:g_epsilon}
\end{equation}
\begin{equation}
f(\varepsilon)\approx\frac{1}{n_{\small{\cal D}}}\sum_{i\in\mathcal{D}}\varepsilon_{i}.
\end{equation}

Assuming $\varepsilon_{i}$, and subsequently $g(\varepsilon)$ and
$f(\varepsilon)$ are small, then Eq.~(\ref{eq:dynamics_dense_theta})
can be expressed as
\begin{eqnarray}
\dot{\varepsilon}_{i} & = & K\sin\alpha+K\left(1-g(\varepsilon)\right)\sin\left(f(\varepsilon)-\varepsilon_{i}-\alpha\right)\nonumber \\
 &  & +\frac{K}{n_{\small{\cal D}}}\rho_{i}\sin(\varphi_{i}-\Omega t-\varepsilon_{i}-\alpha)\nonumber \\
 & \approx & (K\cos\alpha)\left[f(\varepsilon)-\varepsilon_{i}\right]-\left[K\left.\partial_{\varepsilon_{i}}g(\varepsilon)\right|_{\varepsilon_{i}=0}\sin\alpha\right]\varepsilon_{i}\nonumber \\
 &  & +\frac{K}{n_{\small{\cal D}}}\rho_{i}\sin(\varphi_{i}-\Omega t-\alpha)\nonumber \\
 &  & -\frac{K}{n_{\small{\cal D}}}\rho_{i}\cos(\varphi_{i}-\Omega t-\alpha)\varepsilon_{i},\label{eq:epsilon_full}
\end{eqnarray}
where only first order terms in $\varepsilon_{i}$ are retained. Furthermore, in the large
size limit $1/n_{\small{\cal D}}$ is considered to be small and the last term of Eq.~(\ref{eq:epsilon_full})
can be ignored since it contains the product of two small quantities,
$\varepsilon_{i}$ and $1/n_{\small{\cal D}}$. Assuming the fluctuations $\varepsilon_{i}$ are
uncorrelated across sites, $f(\varepsilon)=1/n_{\small{\cal D}}\sum_{i\in\mathcal{D}}\varepsilon_{i}$
smooths out the fluctuations of individual variables and
becomes smaller compared to any single fluctuations $\varepsilon_{i}$; it is therefore
omitted in the following. Additionally, to leading order we have $\left.\partial_{\varepsilon_{i}}g(\varepsilon)\right|_{\varepsilon_{i}=0}=-1/n_{\small{\cal D}}^{2}\sum_{j\neq i}\varepsilon_{j}$,
which vanishes for large $n_{\small{\cal D}}$, such that the corresponding term in Eq.~(\ref{eq:epsilon_full})
can be omitted for fixed $\alpha<\pi/2$, leading to
\begin{equation}
\dot{\varepsilon}_{i}=-(K\cos\alpha)\varepsilon_{i}+\frac{K}{n_{\small{\cal D}}}\rho_{i}\sin(\varphi_{i}-\Omega t-\alpha).\label{eq:epsilon_simp}
\end{equation}

\begin{figure}
\includegraphics[scale=0.4]{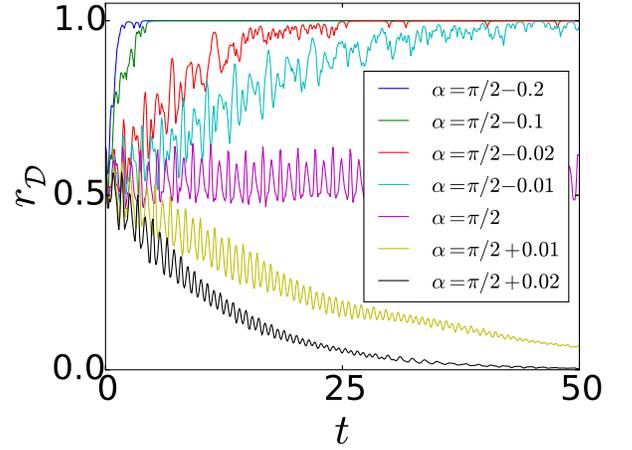}

\caption{(Color online) Evolution of $r_{\small{\cal D}}$ for different phase lags $\alpha$
and a network with $N=200$ variables, sparse degree $d_{\small{\cal S}}=6$ and coefficient $K=10$.\label{fig:stability_demo}}

\end{figure}

By definition of the local mean field of $\rho_{i}$ and $\varphi_{i}$,
the last term of Eq.~(\ref{eq:epsilon_simp}) can be expressed as
\[
\frac{K}{n_{\small{\cal D}}}\rho_{i}\sin(\varphi_{i}-\Omega t-\alpha)=\frac{K}{n_{\small{\cal D}}}\sum_{j\in\partial i\cap\mathcal{S}}\sin(\theta_{j}-\Omega t-\alpha).
\]
Since the phase velocity $|\dot{\theta}_{j}|(\leq Kd_{\small{\cal S}}/n_{\small{\cal D}})$
of the sparsely-connected variables $j$ is much smaller than the oscillation of
$\Omega t$, as shown in Fig.~\ref{fig:dynamic_demo}(a), the two time scales within the $\sin$ function can be separated, allowing one to view $\rho_{i}$ and $\varphi_{i}$ as quench variables within a time frame of a few periods. As
$\Omega=-K(n_{\small{\cal D}}-1)/n_{\small{\cal D}}\sin(\alpha)\approx-K\sin(\alpha)$ for large $n_{\small{\cal D}}$, Eq.~(\ref{eq:epsilon_simp}) can be integrated to provide
\begin{eqnarray}
\varepsilon_{i}(t) & = & c_{1}e^{-(K\cos\alpha)t}+\frac{1}{n_{\small{\cal D}}}\rho_{i}\sin(\varphi_{i}-\Omega t-2\alpha).\label{eq:epsilon_solution}
\end{eqnarray}
This solution Eq. (\ref{eq:epsilon_solution}) has a decay term of rate $K\cos(\alpha)$ and a
stationary oscillatory term of frequency $\Omega=-K\sin(\alpha)$.
The stationary part of $\varepsilon_{i}$ scales as $1/n_{\small{\cal D}}$, which
is consistent with the derivation that treats both $\varepsilon_{i}$
and $1/n_{\small{\cal D}}$ as small quantities at the same time. We can therefore
deduce that the dense group converges faster to the steady solution
for smaller $\alpha$ values, where $\alpha=\pi/2$ signals the onset of instability
of the chimera-like state, as shown in Fig~\ref{fig:stability_demo}.

The deviation $\varepsilon_{i}(t)$ has an oscillatory behavior of frequency
$\Omega$. From Eq.~(\ref{eq:g_epsilon}), we expect that $r_{\small{\cal D}}(t)=1-g(\varepsilon)$
oscillates primarily at a frequency $2\Omega$, which is verified by the
power spectrum of $r_{\small{\cal D}}(t)$ from the numerical experiment in the
inset of Fig.~\ref{fig:dynamic_demo}(b).

\begin{figure}
\includegraphics[scale=0.195]{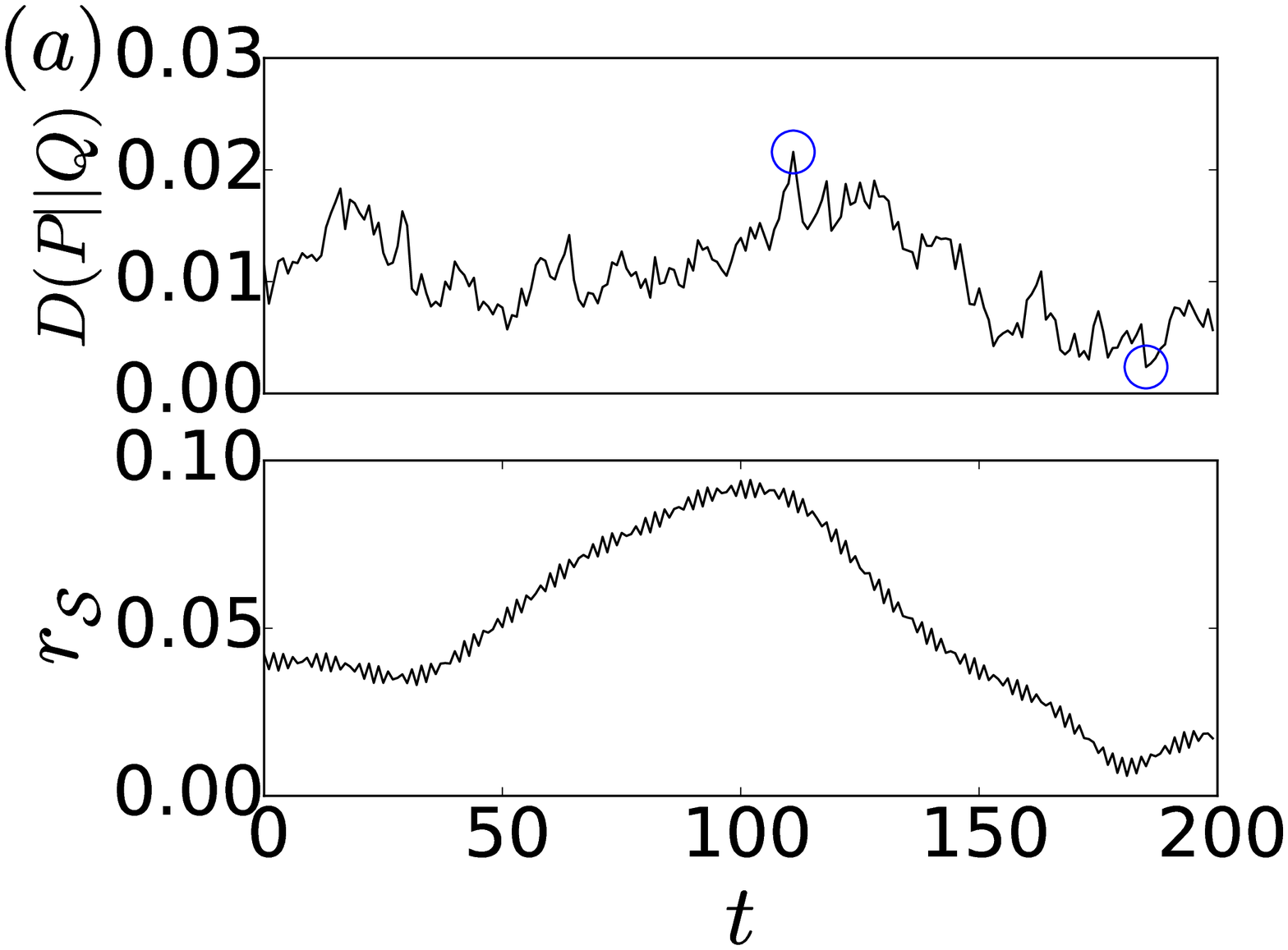}
\includegraphics[scale=0.195]{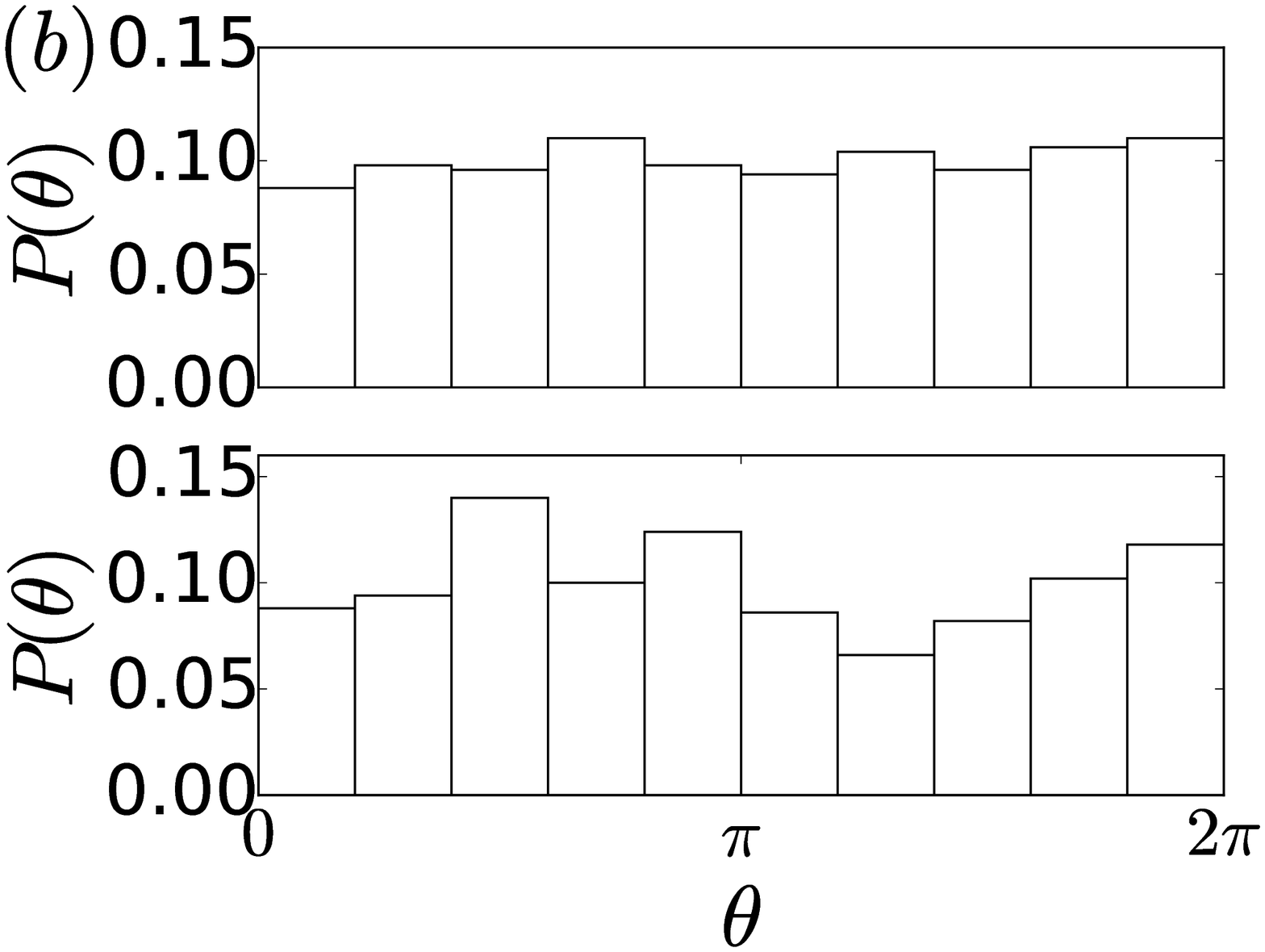}

\caption{(Color online) (a) Top: Kullback\textendash Leibler (KL) divergence
between the phase distribution $P(\theta_{j})$ of the sparse variables
and the uniform distribution $Q(\theta_{j})=const$, defined as $D(P||Q):=\sum_{x}P(x)\ln(P(x)/Q(x))$.
The parameters of the network are $N=1000$, $d_{\small{\cal S}}=6$, $\alpha=\pi/2-0.2$.
$P(\theta_{j})$ is approximated by a histogram with 10 bins, in which
case the upper bound for the KL divergence is $\max_{P}D(P||Q)=\ln10\approx2.30$.
The observed small KL divergence $D(P||Q)$ implies that the phase distribution is close
to uniform. Bottom: The global order parameter $r_{\small{\cal S}}$ for the sparse
group exhibits small values; $r_{\small{\cal S}}=0$ when the phases are uniformly distributed. (b) The distribution $P(\theta_{j})$
at the points with the lowest (upper panel) and highest
(lower panel) KL divergence in the observed time window, as marked
by the two blue circles in the upper panel of (a).\label{fig:sparse_dist}}
\end{figure}

\begin{figure}
\includegraphics[scale=0.2]{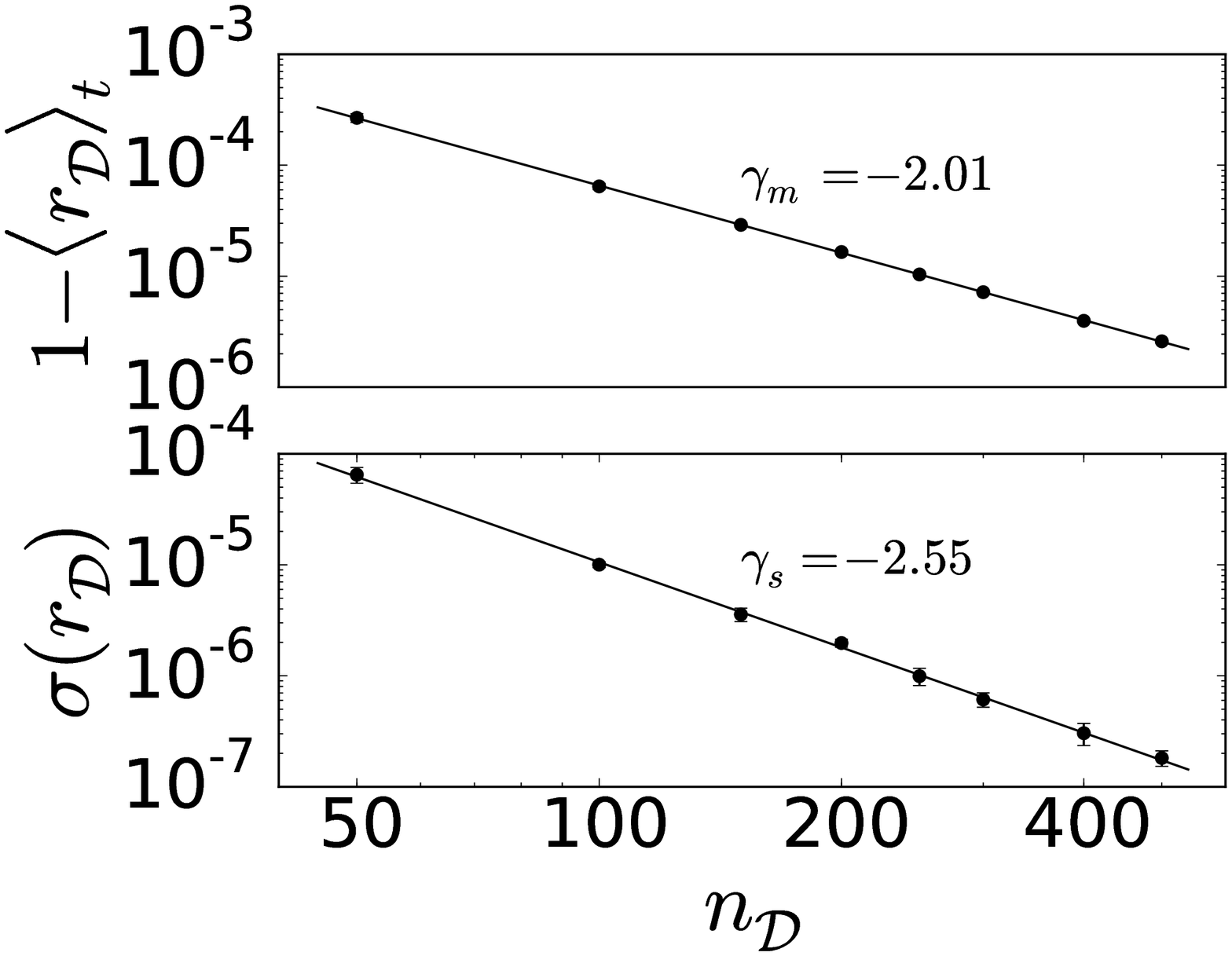}
\includegraphics[scale=0.2]{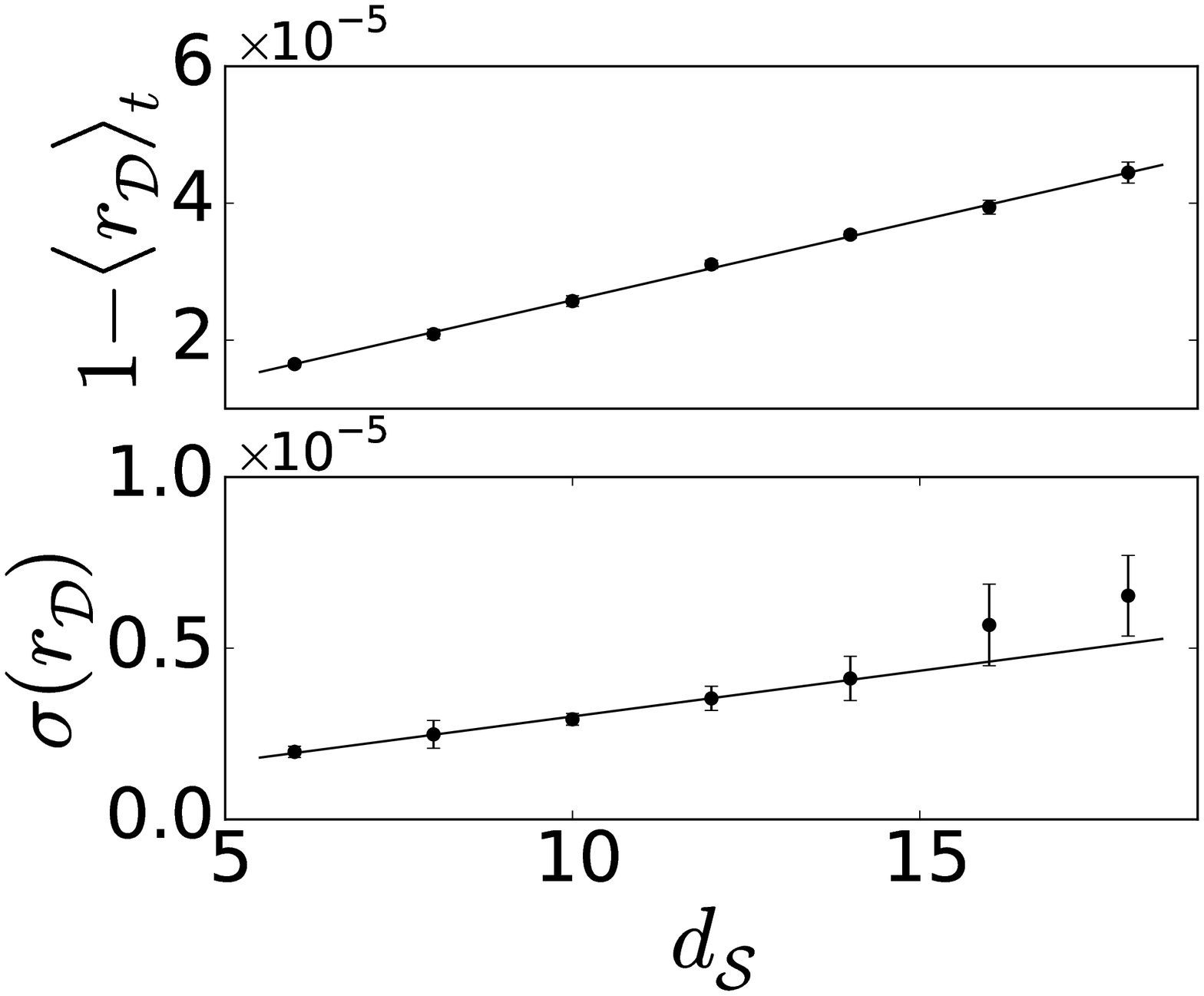}

\caption{(a) Mean and variance of $r_{\small{\cal D}}$ for an observed time window as a
function of $n_{\small{\cal D}}$ with fixed $d_{\small{\cal S}}=6$. The data follows the
scaling of the form $\left(1-\langle r_{\small{\cal D}}\rangle_{t}\right)\sim n_{\small{\cal D}}^{\gamma_{m}}$,
$\sigma(r_{\small{\cal D}})\sim n_{\small{\cal D}}^{\gamma_{s}}$, with coefficients $\gamma_{m}=-2.01$,
$\gamma_{s}=-2.55$. Each data point is averaged over 10 realizations.
(b) The values $1-\langle r_{\small{\cal D}}\rangle_{t}$ and $\sigma(r_{\small{\cal D}})$ as a function
of the sparse degree connectivity $d_{\small{\cal S}}$ with fixed $n_{\small{\cal D}}=200$; both increase linearly
with $d_{\small{\cal S}}$. Each data point is average over 10 realizations. In
the lower panel, the straight line is fitted to the first five data
points, while the measurement of the last two points are rather imprecise and one expects the linear dependence to gradually break down as the ratio $d_{\small{\cal S}}/d_{\small{\cal D}}$ increases.
\label{fig:size_scaling}}
\end{figure}

In the steady state, substituting the solution of $\varepsilon_{i}(t)=\rho_{i}/n_{\small{\cal D}}\sin(\varphi_{i}-\Omega t-2\alpha)$
into Eq.~(\ref{eq:g_epsilon}), one obtains an expression for $g(\varepsilon)$, and
consequently for $r_{\small{\cal D}}$. It is observed that the distribution of sparsely-connected variables phases
is close to uniform, as illustrated
in Fig.~\ref{fig:sparse_dist}. Thus we compute the time average $r_{\small{\cal D}}$
by averaging over the phase angles of the sparse variables, where phases are assumed to be independently and uniformly distributed in
$[0,2\pi)$, i.e., $P(\theta_{j})=1/2\pi$, $\forall j\in\mathcal{S}$.
It is sufficient to consider a single instance of $\varepsilon_{i}(t)$
\[
\epsilon_{i}=\frac{\rho_{i}}{n_{\small{\cal D}}}\sin(\varphi_{i})=\frac{1}{n_{\small{\cal D}}}\sum_{j\in\partial i\text{\ensuremath{\cap}}\mathcal{S}}\sin(\theta_{j}).
\]
Since $\mathbb{E}[\sin(\theta_{j})]=0$, $\mathbb{E}[\sin^{2}(\theta_{j})]=1/2$,
$\mathbb{E}[\sin^{4}(\theta_{j})]=3/8$, it is straightforward to
obtain the mean and variance of $r_{\small{\cal D}}$,
\begin{eqnarray}
\mathbb{E}[r_{\small{\cal D}}] & = & \mathbb{E}\left[1-g(\epsilon)\right]\nonumber \\
 & = & 1-\frac{1}{2n_{\small{\cal D}}}\mathbb{E}\left[\sum_{i\in\mathcal{D}}\epsilon_{i}^{2}\right]+\frac{1}{2n_{\small{\cal D}}^{2}}\mathbb{E}\left[\left(\sum_{i\in\mathcal{D}}\epsilon_{i}\right)^{2}\right]\\
 & = & 1-\frac{d_{\small{\cal S}}}{8n_{\small{\cal D}}^{2}}+\frac{d_{\small{\cal S}}^{2}}{16n_{\small{\cal D}}^{3}},\label{eq:r1_mean}
\end{eqnarray}

\begin{eqnarray}
\mathbb{E}[r_{\small{\cal D}}^{2}]-\mathbb{E}[r_{\small{\cal D}}]^{2} & = & \mathbb{E}\left[\left(1-g(\epsilon)\right)^{2}\right]-\mathbb{E}[1-g(\epsilon)]^{2}\nonumber \\
 & = & \frac{1}{32n_{\small{\cal D}}^{5}}\left(d_{\small{\cal S}}^{2}-\frac{3}{2}d_{\small{\cal S}}\right)+O\left(\frac{1}{n_{\small{\cal D}}^{6}}\right).\label{eq:r1_variance}
\end{eqnarray}
Therefore, to leading order, the mean of $1-\mathbb{E}[r_{\small{\cal D}}]$ scales
as $d_{\small{\cal S}}/n_{\small{\cal D}}^{2}$, while the standard deviation $\sigma(r_{\small{\cal D}})$
scales as $d_{\small{\cal S}}/n_{\small{\cal D}}^{2.5}$. The finite size scaling behavior is
confirmed, in full agreement, by numerical simulations as shown in Fig.~\ref{fig:size_scaling}.

\begin{figure}

\includegraphics[scale=0.4]{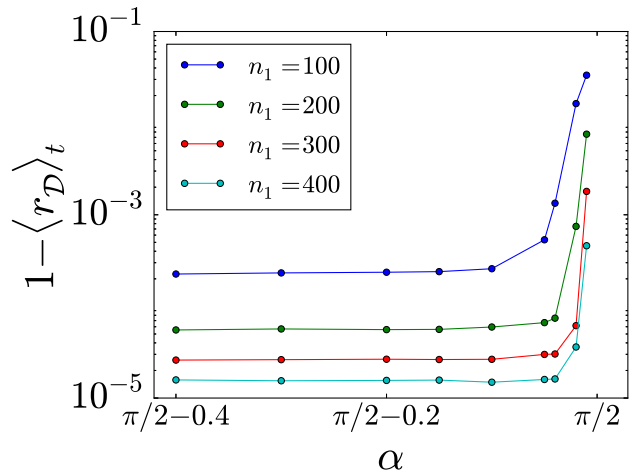}

\caption{Expected first order corrections $1-\text{\ensuremath{\langle}}r_{\small{\cal D}}\rangle_{t}$ vs $\alpha$ for
different system sizes, measured at the stationary states.\label{fig:insensitive_to_alpha} }
\end{figure}

The obtained solution of $r_{\small{\cal D}}$ in Eq.~(\ref{eq:r1_mean}) is independent
of $\alpha$. Nevertheless, we remark that the derivation breaks down
when $\alpha$ is close to $\pi/2$, in particular for $\alpha\sim\tan^{-1}(n_{\small{\cal D}}^{2})$,
where the term $\left[K\left.\partial_{\varepsilon_{i}}g(\varepsilon)\right|_{\varepsilon_{i}=0}\sin\alpha\right]\varepsilon_{i}$
starts to dominate $(K\cos\alpha)\varepsilon_{i}$ in Eq.~(\ref{eq:epsilon_full}).
We can see this effect in Fig.~\ref{fig:insensitive_to_alpha} where
for a wide range of $\alpha$ values away from $\pi/2$, the deviation to
perfect synchronization in the stationary state is independent of
$\alpha$. When $\alpha$ gets closer to $\pi/2$, the deviation
starts to increase; the onset of this deviation occurs later for larger systems, as
expected.

\section{Discussion and  Conclusion}

We study a simple heterogeneous network architecture that supports easy-to-reach chimera-like states and is amenable to analysis. 
The network comprises two interconnected components of densely connected and sparsely connected variables. 
The heterogeneity separates the time scales of the dynamics of the two components, leading to disparate behaviors of the two groups. We showed that the densely connected group maintains a synchronized and coherent dynamics, while the dynamics of sparse group variables is incoherent. 
Furthermore, we derived a self-consistent mean field theory for these networks, by viewing the contribution of the sparse variables as $1/n_{\small{\cal D}}$-scale perturbations to the dense group's dynamics. 
Using this framework we obtained expressions for the dense group order parameter that exhibits phase synchronization, and subsequently its dependence on phase lag parameter, network size and network connectivity. 
The model facilitates the analysis of large systems comprising interacting oscillatory variables to provide new insight and understanding of their complex dynamical behaviors. 

In general, the densely connected nodes in the complex network may not form a regular subgraph. 
One possible generalization of our model is to consider randomly removal of connections in the dense group~\citep{Laing2012}, and study the resulting erosion of synchronization in the framework developed in related works~\citep{Skardal2015, Skardal2016}.
In more complex heterogeneous network topologies such as scale free networks, although our analysis is not directly applicable, we suggest that similar mechanism accounts for the disparate synchronization behaviors for the high-degree nodes and other nodes, i.e., the high-degree nodes interact with each other strongly and in a combinatorial sense they have higher chance to arrange themselves to a synchronized cluster, while the incoherent perturbations from the low-degree nodes are less significant, as preliminarily illustrated in numerical simulations~\citep{Zhu2014}.  
We envisage that our model and similar variants could be employed to explore the properties of other heterogeneous network architectures.

\section*{Acknowledgement}
Support from the Overseas Research Awards of Hong Kong University of Science and Technology (BL) and  The Leverhulme Trust grant RPG-2013-48 (DS) is acknowledged. The work was carried out during an extended visit to Aston University, Bo Li would like to thank Aston for the hospitality.

\appendix
\section{The Expressions of $g(\varepsilon)$ and $f(\varepsilon)$}\label{appendix_a}
In this section we derive the expressions of $g(\varepsilon)$ and $f(\varepsilon)$ up to leading order of $\varepsilon^2$. From Eq.~(\ref{eq:r_D}), we see that $r_{\mathcal{D}}=1-g(\varepsilon)$ is the magnitude of the complex vector $\frac{1}{n_{\small{\cal D}}}\sum_{i\in\mathcal{D}}e^{i\varepsilon_{i}} \approx 1-\frac{1}{2n_{\small{\cal D}}}\sum_{i\in\mathcal{D}}\varepsilon_{i}^{2} +\frac{i}{n_{\small{\cal D}}}\sum_{i\in\mathcal{D}}\varepsilon_{i}$, then
\begin{eqnarray*}
r_{\mathcal{D}} & \approx & \sqrt{\left(1-\frac{1}{2n_{\mathcal{D}}}\sum_{i\in\mathcal{D}}\varepsilon_{i}^{2}\right)^{2}+\left(\frac{1}{n_{\mathcal{D}}}\sum_{i\in\mathcal{D}}\varepsilon_{i}\right)^{2}}\\
 & \approx & \sqrt{1-\frac{1}{n_{\mathcal{D}}}\sum_{i\in\mathcal{D}}\varepsilon_{i}^{2}+\left(\frac{1}{n_{\mathcal{D}}}\sum_{i\in\mathcal{D}}\varepsilon_{i}\right)^{2}}\\
 & \approx & 1-\frac{1}{2n_{\mathcal{D}}}\sum_{i\in\mathcal{D}}\varepsilon_{i}^{2}+\frac{1}{2n_{\mathcal{D}}^{2}}\left(\sum_{i\in\mathcal{D}}\varepsilon_{i}\right)^{2},
\end{eqnarray*}
where we have used the approximation $\sqrt{1+x} \approx 1+x/2$ for small $x$. Therefore, we obtain the expression
\[
g(\varepsilon)=\frac{1}{2n_{\mathcal{D}}}\sum_{i\in\mathcal{D}}\varepsilon_{i}^{2}-\frac{1}{2n_{\mathcal{D}}^{2}}\left(\sum_{i\in\mathcal{D}}\varepsilon_{i}\right)^{2}.
\]

Notice that $f(\varepsilon)$ is the phase of the complex vector $\frac{1}{n_{\small{\cal D}}}\sum_{i\in\mathcal{D}}e^{i\varepsilon_{i}}$; after some manipulation we find that
\begin{eqnarray*}
f(\varepsilon) & = & \tan^{-1}\frac{\frac{1}{n_{\mathcal{D}}}\sum_{i\in\mathcal{D}}\varepsilon_{i}}{1-\frac{1}{2n_{\mathcal{D}}}\sum_{i\in\mathcal{D}}\varepsilon_{i}^{2}}\\
 & \approx & \frac{1}{n_{\mathcal{D}}}\sum_{i\in\mathcal{D}}\varepsilon_{i}+\frac{1}{2n_{\mathcal{D}}^{2}}\sum_{i\in\mathcal{D}}\varepsilon_{i}\sum_{i\in\mathcal{D}}\varepsilon_{i}^{2}.
\end{eqnarray*}

\bibliographystyle{unsrt}
\bibliography{ref}

\begin{thebibliography}{10}

\bibitem{Acebron2005}
Juan~A. Acebr\'on, L.~L. Bonilla, Conrad~J. P\'erez~Vicente, F\'elix Ritort,
  and Renato Spigler.
\newblock The {Kuramoto} model: A simple paradigm for synchronization
  phenomena.
\newblock {\em Rev. Mod. Phys.}, 77:137--185, Apr 2005.

\bibitem{Boccaletti2002}
S.~Boccaletti, J.~Kurths, G.~Osipov, D.L. Valladares, and C.S. Zhou.
\newblock The synchronization of chaotic systems.
\newblock {\em Physics Reports}, 366(1-2):1 -- 101, 2002.

\bibitem{Rodrigues2016}
Francisco~A. Rodrigues, Thomas K.~DM. Peron, Peng Ji, and J{\"u}rgen Kurths.
\newblock The {Kuramoto} model in complex networks.
\newblock {\em Physics Reports}, 610:1 -- 98, 2016.

\bibitem{Panaggio2015}
Mark~J Panaggio and Daniel~M Abrams.
\newblock Chimera states: coexistence of coherence and incoherence in networks
  of coupled oscillators.
\newblock {\em Nonlinearity}, 28(3):R67, 2015.

\bibitem{Zhu2014}
Yun Zhu, Zhigang Zheng, and Junzhong Yang.
\newblock Chimera states on complex networks.
\newblock {\em Phys. Rev. E}, 89:022914, Feb 2014.

\bibitem{Kuramoto2002}
Kuramoto Y. and Battogtokh D.
\newblock Coexistence of coherence and incoherence in nonlocally coupled phase
  oscillators.
\newblock {\em Nonlinear Phenomena In Complex Systems}, 5(4):380--385, 2002.

\bibitem{Shima2004}
Shin-ichiro Shima and Yoshiki Kuramoto.
\newblock Rotating spiral waves with phase-randomized core in nonlocally
  coupled oscillators.
\newblock {\em Phys. Rev. E}, 69:036213, Mar 2004.

\bibitem{Abrams2004}
Daniel~M. Abrams and Steven~H. Strogatz.
\newblock Chimera states for coupled oscillators.
\newblock {\em Phys. Rev. Lett.}, 93:174102, Oct 2004.

\bibitem{Abrams2008}
Daniel~M. Abrams, Rennie Mirollo, Steven~H. Strogatz, and Daniel~A. Wiley.
\newblock Solvable model for chimera states of coupled oscillators.
\newblock {\em Phys. Rev. Lett.}, 101:084103, Aug 2008.

\bibitem{Ashwin2015}
Peter Ashwin and Oleksandr Burylko.
\newblock Weak chimeras in minimal networks of coupled phase oscillators.
\newblock {\em Chaos: An Interdisciplinary Journal of Nonlinear Science},
  25(1):013106, 2015.

\bibitem{Bick2017}
Christian Bick.
\newblock Isotropy of angular frequencies and weak chimeras with broken
  symmetry.
\newblock {\em Journal of Nonlinear Science}, 27(2):605--626, 2017.

\bibitem{Saad2013}
David Saad and Alexander Mozeika.
\newblock Emergence of equilibriumlike domains within nonequilibrium ising spin
  systems.
\newblock {\em Phys. Rev. E}, 87:032131, Mar 2013.

\bibitem{Ott2008}
Edward Ott and Thomas~M. Antonsen.
\newblock Low dimensional behavior of large systems of globally coupled
  oscillators.
\newblock {\em Chaos: An Interdisciplinary Journal of Nonlinear Science},
  18(3):037113, 2008.

\bibitem{Martens2016}
Erik~A Martens, Mark~J Panaggio, and Daniel~M Abrams.
\newblock Basins of attraction for chimera states.
\newblock {\em New Journal of Physics}, 18(2):022002, 2016.

\bibitem{Tinsley2012}
Mark~R. Tinsley, Simbarashe Nkomo, and Kenneth Showalter.
\newblock Chimera and phase-cluster states in populations of coupled chemical
  oscillators.
\newblock {\em Nat Phys}, 8(9):662--665, Sep 2012.

\bibitem{Hagerstrom2012}
Aaron~M. Hagerstrom, Thomas~E. Murphy, Rajarshi Roy, Philipp Hovel, Iryna
  Omelchenko, and Eckehard Scholl.
\newblock Experimental observation of chimeras in coupled-map lattices.
\newblock {\em Nat Phys}, 8(9):658--661, Sep 2012.

\bibitem{Martens2013}
Erik~Andreas Martens, Shashi Thutupalli, Antoine Fourrière, and Oskar
  Hallatschek.
\newblock Chimera states in mechanical oscillator networks.
\newblock {\em Proceedings of the National Academy of Sciences},
  110(26):10563--10567, 2013.

\bibitem{Kapitaniak2014}
Tomasz Kapitaniak, Patrycja Kuzma, Jerzy Wojewoda, Krzysztof Czolczynski, and
  Yuri Maistrenko.
\newblock Imperfect chimera states for coupled pendula.
\newblock {\em Scientific Reports}, 4:6379, Sep 2014.
\newblock Article.

\bibitem{Strogatz2001}
Steven~H. Strogatz.
\newblock Exploring complex networks.
\newblock {\em Nature}, 410(6825):268--276, Mar 2001.

\bibitem{Albert2002}
R\'eka Albert and Albert-L\'aszl\'o Barab\'asi.
\newblock Statistical mechanics of complex networks.
\newblock {\em Rev. Mod. Phys.}, 74:47--97, Jan 2002.

\bibitem{Arenas2008}
Alex Arenas, Albert D{\'i}az-Guilera, J{\"u}rgen Kurths, Yamir Moreno, and
  Changsong Zhou.
\newblock Synchronization in complex networks.
\newblock {\em Physics Reports}, 469(3):93 -- 153, 2008.

\bibitem{Vuksanovic2014}
Vesna Vuksanovi{\'c} and Philipp H{\"o}vel.
\newblock Functional connectivity of distant cortical regions: Role of remote
  synchronization and symmetry in interactions.
\newblock {\em NeuroImage}, 97:1 -- 8, 2014.

\bibitem{Hizanidis2016}
Johanne Hizanidis, Nikos~E. Kouvaris, Gorka Zamora-L{\'o}pez, Albert
  D{\'i}az-Guilera, and Chris~G. Antonopoulos.
\newblock Chimera-like states in modular neural networks.
\newblock {\em Scientific Reports}, 6:19845, Jan 2016.
\newblock Article.

\bibitem{Ko2008}
Tae-Wook Ko and G.~Bard Ermentrout.
\newblock Partially locked states in coupled oscillators due to inhomogeneous
  coupling.
\newblock {\em Phys. Rev. E}, 78:016203, Jul 2008.

\bibitem{Laing2009a}
Carlo~R. Laing.
\newblock Chimera states in heterogeneous networks.
\newblock {\em Chaos: An Interdisciplinary Journal of Nonlinear Science},
  19(1):013113, 2009.

\bibitem{Laing2009b}
Carlo~R. Laing.
\newblock The dynamics of chimera states in heterogeneous kuramoto networks.
\newblock {\em Physica D: Nonlinear Phenomena}, 238(16):1569 -- 1588, 2009.

\bibitem{Laing2012}
Carlo~R. Laing, Karthikeyan Rajendran, and Ioannis~G. Kevrekidis.
\newblock Chimeras in random non-complete networks of phase oscillators.
\newblock {\em Chaos: An Interdisciplinary Journal of Nonlinear Science},
  22(1):013132, 2012.

\bibitem{Omelchenko2015}
Iryna Omelchenko, Astero Provata, Johanne Hizanidis, Eckehard Sch\"oll, and
  Philipp H\"ovel.
\newblock Robustness of chimera states for coupled fitzhugh-nagumo oscillators.
\newblock {\em Phys. Rev. E}, 91:022917, Feb 2015.

\bibitem{Buscarino2015}
Arturo Buscarino, Mattia Frasca, Lucia~Valentina Gambuzza, and Philipp H\"ovel.
\newblock Chimera states in time-varying complex networks.
\newblock {\em Phys. Rev. E}, 91:022817, Feb 2015.

\bibitem{Jiang2016}
Xin Jiang and Daniel~M. Abrams.
\newblock Symmetry-broken states on networks of coupled oscillators.
\newblock {\em Phys. Rev. E}, 93:052202, May 2016.

\bibitem{Olmi2016}
Simona Olmi and Alessandro Torcini.
\newblock Chimera states in pulse coupled neural networks: the influence of
  dilution and noise.
\newblock {\em e-print}, arXiv:1606.08618, 2017.

\bibitem{Zhou2006}
Changsong Zhou and J{\"u}rgen Kurths.
\newblock Hierarchical synchronization in complex networks with heterogeneous
  degrees.
\newblock {\em Chaos}, 16(1):015104, 2006.

\bibitem{Gardenes2007}
Jes{\'u}s G{\'o}mez-Garde{\~n}es, Yamir Moreno, and Alex Arenas.
\newblock Paths to synchronization on complex networks.
\newblock {\em Phys. Rev. Lett.}, 98:034101, Jan 2007.

\bibitem{Gardenes2011}
Jes{\'u}s G{\'o}mez-Garde\~nes, Sergio G{\'o}mez, Alex Arenas, and Yamir
  Moreno.
\newblock Explosive synchronization transitions in scale-free networks.
\newblock {\em Phys. Rev. Lett.}, 106:128701, Mar 2011.

\bibitem{Pereira2010}
Tiago Pereira.
\newblock Hub synchronization in scale-free networks.
\newblock {\em Phys. Rev. E}, 82:036201, Sep 2010.

\bibitem{Eckmann2010}
Jean-Pierre Eckmann, Elisha Moses, Olav Stetter, Tsvi Tlusty, and Cyrille
  Zbinden.
\newblock Leaders of neuronal cultures in a quorum percolation model.
\newblock {\em Frontiers in Computational Neuroscience}, 4:132, 2010.

\bibitem{Skardal2015}
Per~Sebastian Skardal, Dane Taylor, Jie Sun, and Alex Arenas.
\newblock Erosion of synchronization in networks of coupled oscillators.
\newblock {\em Phys. Rev. E}, 91:010802, Jan 2015.

\bibitem{kuramoto1984}
Y.~Kuramoto.
\newblock {\em Chemical Oscillations, Waves, and Turbulence}.
\newblock Springer-Verlag Berlin Heidelberg, 1984.

\bibitem{Skardal2016}
Per~Sebastian Skardal, Dane Taylor, Jie Sun, and Alex Arenas.
\newblock Erosion of synchronization: Coupling heterogeneity and network
  structure.
\newblock {\em Physica D: Nonlinear Phenomena}, 323–324:40 -- 48, 2016.

\end{thebibliography}

\end{document}